\documentclass[twocolumn,superscriptaddress,amsmath,amssymb]{revtex4}

\usepackage{graphicx}
\usepackage{dcolumn}
\usepackage{bm}

\begin{document}

\title{Metal-nonmetal transition in Li$_x$CoO$_2$ thin film 
and thermopower enhancement at high Li concentration}

\author{Y.~Ishida}
\affiliation{RIKEN SPring-8 Center, Sayo, Sayo, Hyogo 679-5148, 
Japan}

\author{A.~Mizutani}
\author{K.~Sugiura}
\author{H.~Ohta}
\affiliation{Graduate School of Engineering, Nagoya University, 
Furo-cho, Chikusa, Nagoya 464-8603, Japan}

\author{K.~Koumoto}
\affiliation{Graduate School of Engineering, Nagoya University, 
Furo-cho, Chikusa, Nagoya 464-8603, Japan}
\affiliation{CREST, Japan Science and Technology Agency,
3 Ban-cho, Chiyoda, Tokyo 102-0075, Japan
}

\begin{abstract}
We investigate the transport properties of Li$_x$CoO$_2$ 
thin films whose resistivities are nearly an order 
of magnitude lower than those of the bulk polycrystals. 
A metal-nonmetal transition occurs at $x\sim$ 0.8 in a biphasic 
domain, and the Seebeck coefficient ($S$) 
is drastically 
increased at $\sim$140 K ($\equiv T^*$) 
with increasing the Li concentration to show a 
peak of magnitude $\sim$120 $\mu$V/K in the $S$-$T$ curve of 
$x =$ 0.87. 
We show that $T^*$ corresponds to a crossover temperature 
in the conduction, most likely reflecting the correlation-induced 
temperature dependence in the low-energy 
excitations. 
\end{abstract}

\maketitle

Thermoelectric (TE) energy conversion is one of the key 
technologies for energy savings and environmental protection. 
There has been an extensive quest over decades for materials exhibiting 
higher TE figure of merit $ZT = S^2T$/$\rho\kappa$, 
where $S$, $T$, $\rho$, and $\kappa$ are the Seebeck coefficient, 
absolute temperature, resistivity, and thermal conductivity, 
respectively \cite{Mahan_Today}. 
Band theory of solids predicts that heavily-doped semiconductors 
having carrier concentrations $n\sim$ 10$^{19}$ cm$^{-3}$ 
are candidates for high-$ZT$ materials, but 
metallic materials are not \cite{Mahan_Today}. 
Therefore, the fairly large $S$ ($\sim$100 $\mu$V/K at 300 K) 
in Na$_x$CoO$_2$ showing metallic $\rho$ with 
$n\gtrsim$ 10$^{21}$\,cm$^{-3}$ 
\cite{Terasaki} 
has attracted much attention 
for its origin \cite{Singh, Koshibae_PRB, WangOng, PALee, 
Haerter, Limelette, Singh2007, Kuroki_JPSJ, 
Ishida_JPSJ} and also has triggered the search for metallic TE materials 
having CoO$_2$ layers as common building blocks 
\cite{Kajitani, Lee_NMat, Kanno_Sr, Sugiura_Sr, Masset_Ca, 
Funahashi_Ca, Sugiura_Ca}. 
Na$_x$CoO$_2$ was further revealed to exhibit 
unconventional superconductivity \cite{Super}, charge \cite{MLFoo} 
and Na-ion orderings \cite{Roger}, 
and three-dimensonal magnetism \cite{Sugiyama_PRB, Bayrakci}. 
Electronically, 
$A$-ion removal from $A_x$CoO$_2$ 
($A$: alkaline or alkaline-earth metal) creates 
holes in the valence band. It has been debated 
how far band theory can explain the largeness of $S$ 
\cite{Singh, Singh2007, Kuroki_JPSJ} 
and how far strong correlations can be relevant 
\cite{Koshibae_PRB, WangOng, PALee, Haerter, Limelette, Ishida_JPSJ}. 
Recently, Lee {\it et al}.\ \cite{Lee_NMat} 
reported that $S$ is further increased 
in Na$_x$CoO$_2$ at $x >$ 0.75, 
and that the $S$-$T$ curve shows a 
peculiar peak of magnitude $\sim$350 $\mu$V/K at $\sim$100 K, 
indicating that Na$_x$CoO$_2$ in the high-$x$ region 
may serve as an efficient hole-type TE material, although the origin 
of the increase in $S$ at high $x$ 
is also not clear partly because Na ions are unstable and mobile 
at ambient temperatures \cite{Roger, Batlogg, Julien}. 

Li$_x$CoO$_2$ \cite{Li_1, Mendiboure} has a 
layered structure similar to Na$_x$CoO$_2$, 
and is the most common cathode material for Li rechargeable 
batteries since 
Li ions can be removed from and 
inserted into Li$_x$CoO$_2$ repeatedly 
through an electrochemical method at ambient temperatures 
\cite{Mizushima, LiNMR}. 
There has been many studies on the relationship between the 
structure and the Li-ion diffusion properties as reviewed by 
Antolini \cite{Antolini}, and also on the magnetic properties 
\cite{Vaulx_PRL, Mukai_PRL, Cava_PRB, Delmas2009, Motohashi_PRB}. 
In addition, 
Li$_x$CoO$_2$ is known to exhibit fairly large $S$ 
\cite{Honders, Molenda, Menetrier} comparable to  
that of Na$_x$CoO$_2$. Even though $\rho$ of the polycrystalline 
samples are rather high due to the grain boundaries, 
there is a signature of electron delocalization upon Li deintercalation 
\cite{Molenda, Menetrier}, 
indicating that Li$_x$CoO$_2$ is a potentially metallic 
TE material. 
Herein, we report transport properties of 
Li$_x$CoO$_2$ epitaxial thin films 
whose resistivities are 
nearly an order of magnitude lower than those of the 
polycrystalline samples. 
We observe a metal-nonmetal transition at $x\sim$ 0.8 
and an increase in $S$ with increasing $x$ that results into 
a peak in the $S$-$T$ curve at high Li concentration. 
We present a picture that the electronic transport 
of Li$_x$CoO$_2$ at high $x$ is strongly affected 
by both disorder and electron correlations.

\begin{figure}[htb]
\begin{center}
\includegraphics[width = 7.5 cm]{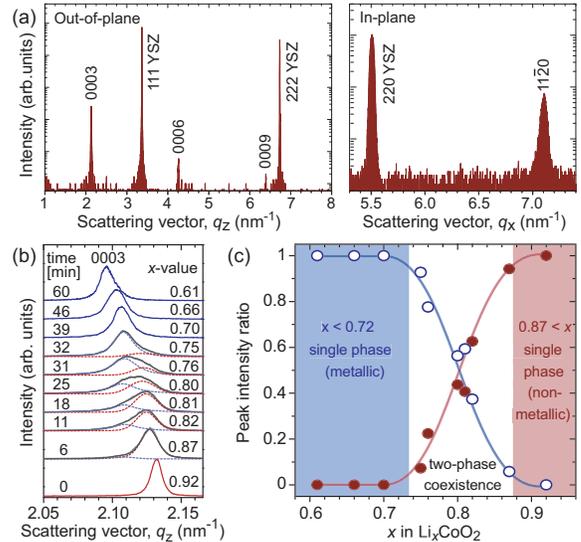}
\caption{\label{fig1} 
XRD patterns of Li$_x$CoO$_2$ thin films. 
(a) Out-of-plane (left) and in-plane (right) XRD patterns of 
Li$_{0.92}$CoO$_2$ epitaxial thin film. 
(b) 0003 peaks of Li$_x$CoO$_2$ thin films. With increasing the immersing 
time, the $z$ component of the scattering vector is decreased, from which 
we deduced the Li concentration $x$. Phase separation is observed 
for the samples 
0.75 $\le$ $x$ $\le$ 0.87, in which the 0003 peak can be deconvoluted 
into two components [red (gray) dashed curve: a peak originating from the 
high Li concentration phase; blue (light gray) dashed curve: 
a peak originating from the low Li concentration phase]. 
(c) Ratios of the high (filled circle) and the low (empty circle) 
Li concentration 
phases deduced from the 0003 peak analysis shown in (b).
}
\end{center}
\end{figure}

In order to obtain the Li$_x$CoO$_2$ epitaxial thin films, 
we first prepared 
Na$_{0.8}$CoO$_2$/(111)YSZ by the reactive 
solid-phase epitaxy method \cite{Mizutani, Ohta_AdFunc, Ohta_CGD}. 
The Na$_{0.8}$CoO$_2$ epitaxial thin films thus obtained were 
converted into Li$_{0.92}$CoO$_2$ thin films 
by treating them in LiNO$_3$/LiCl 
salt at 260$^{\circ}$C for 4 h in an Ar atmosphere. 
During the topotactic ion-exchange treatment, 
the epitaxial relation of the film and the YSZ substrate 
was preserved 
[the in-plane and out-of-plane XRD patterns 
show sharp peaks of the films and the substrate simultaneously; 
see Fig.\ \ref{fig1}(a)], 
and the grain size of $\sim$0.5 $\mu$m was virtually unchanged, 
thanks to the lateral ion diffusion during the ion exchange 
making less impact on the crystallinity of the CoO$_2$ layers 
\cite{Mizutani, Ohta_AdFunc, Ohta_CGD}. 
Then the Li ions were subtracted 
by immersing the Li$_{0.92}$CoO$_2$ thin film 
in K$_2$S$_2$O$_8$ 
aqueous solution (0.03 mol/$l$) at room temperature. 
The Li concentrations were controlled by the immersing time, 
and $x$ was estimated from the 
0003 XRD peak [Fig.\ \ref{fig1}(b)] 
using the relationship between $x$ and 
the lattice parameter of the bulk \cite{Mizutani, Reimers, foot}. 
The thin films for 0.75 $\le x \le$ 0.87 was separated 
into Li-rich and Li-poor 
phases, whereas those for 
$x \le$ 0.70 and 0.92 $\le x$ were in a 
single phase. 
A two-phase coexistence was also reported 
in the polycrystalline samples at 0.75 $\le x \le$ 0.95
\cite{Reimers, Antolini}, 
so that the thin films investigated herein had 
structural properties similar to the bulk. 
$\rho$ was recorded by a d.c.\ 
four probe method under van der Pauw configuration 
and $S$ was recorded by 
applying temperature difference of $\textless$3 K in the 
in-plane direction. 
Temperature-dependent hysteresis was not observed in the 
transport measurements. 
Typical thickness of the films was $\sim$60 nm. 

\begin{figure}[htb]
\begin{center}
\includegraphics[width = 8.0 cm]{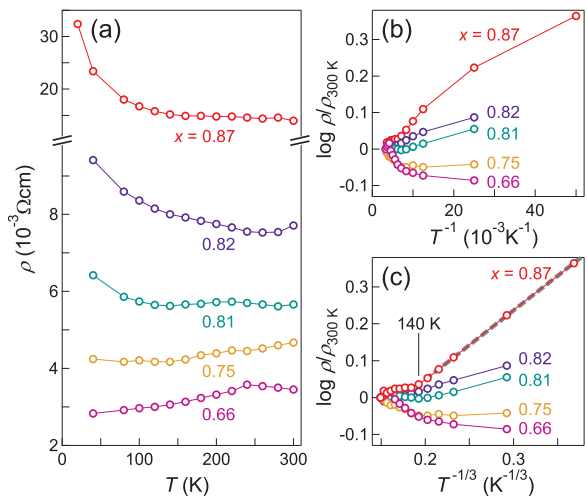}
\caption{\label{fig2} 
Temperature dependence of $\rho$ of 
the Li$_x$CoO$_2$ epitaxial films. (a) A $\rho$ - $T$ plot. 
(b) A log $\rho$/$\rho_{\rm 300 K}$-$T^{-1}$ plot for 
evaluating the activated-type conduction. 
(c) A log $\rho$/$\rho_{\rm 300 K}$-$T^{-1/3}$ plot 
for evaluating the two-dimensional VRH. The dashed 
line indicates the linear relationship 
between log $\rho$/$\rho_{\rm 300 K}$ and $T^{-1/3}$ at 
$T\lesssim$ 140 K for the $x$ = 0.87 sample. 
}
\end{center}
\end{figure}

Figure \ref{fig2}(a) shows $\rho$-$T$ curves of the Li$_x$CoO$_2$ 
epitaxial films. 
The $x$ = 0.66 sample shows positive 
$\partial\rho$/$\partial T$ at $T\lesssim$ 240 K, 
indicating that the single phase samples 
for $x \le$ 0.70 is metallic at low $T$. 
With increasing $x$, $\rho$ is increased, 
and a metal-nonmetal transition occurs at 
$x\sim$ 0.8 ($\equiv x_c$) evidenced by a change in sign 
in $\partial\rho$/$\partial T$ 
as reported previously \cite{Molenda, Menetrier}. 
However, $\rho$ reported 
herein is nearly an order of magnitude lower than those of the 
polycrystalline samples presumably due to the 
reduced grain-boundary effects since the CoO$_2$ layers in the 
planer grains are connected laterally in the epitaxial thin films. 
We also note that there is little anomaly in 
$\rho$ at $\sim$175 K, where the 
magnetic susceptibility 
shows an anomaly 
for the samples having fractional Li contents 
\cite{Motohashi_PRB}. 

\begin{figure}[htb]
\begin{center}
\includegraphics[width = 8.3 cm]{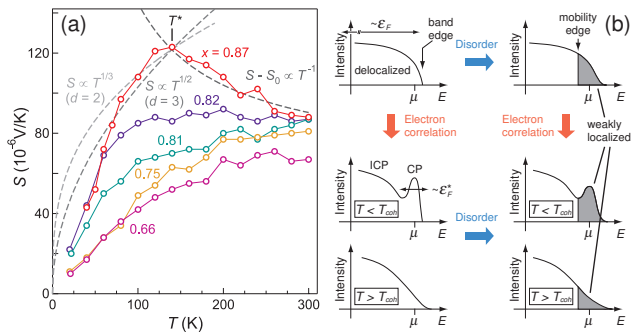}
\caption{\label{fig3} 
Temperature dependence of 
$S$. (a) $S$-$T$ curves and 
theoretical curves for the VRH and the nearest-neighbor hopping (see text). 
(b) Excitation spectra near $\mu$ 
affected by disorder (right top), 
electron correlation (left bottom), and both disorder and correlation 
(right bottom). The spectrum of 
Na$_x$CoO$_2$ exhibits coherent part (CP) and incoherent part (ICP) 
at $T < T_{\it coh}$ due to correlation \cite{Valla, Hasan}.}
\end{center}
\end{figure}

In order to investigate the type of conduction, 
we plot $\log \rho/\rho_{\rm 300 K}$\,-\,$T^{-1}$ 
and $\log \rho/\rho_{\rm 300 K}$\,-\,$T^{-1/3}$ 
in Figs.\ \ref{fig2}(b) and \ref{fig2}(c), 
respectively, where $\rho_{\rm 300 K}$ is the resistivity 
at 300 K. 
The diverging behavior of $\rho$ with $T\,\to\,0$ 
of the $x =$ 0.87 sample is not as strong as that expected in 
an activated-type conduction, 
$\rho \propto \exp(\Delta/k_{B}T)$ [$\Delta$ is the 
energy gap in the density of states (DOS) about the chemical 
potential, $\mu$], 
but is indicative of a variable-range-hopping (VRH) conduction, where 
$\rho \propto \exp(T_0/T)^{1/(d+1)}$ ($d$ is the dimension 
of the hopping conduction \cite{Mott}). 
One can see that the two-dimensional VRH conduction, 
$\rho \propto \exp(T_0/T)^{1/3}$ 
with $T_0$ = 81 K [dashed line in Fig.\ \ref{fig2}(c)] 
can reproduce the $\rho$-$T$ curve of the $x$ = 0.87 
sample fairly well at 
$T \lesssim$ 140 K ($\equiv T^*$), 
thus revealing a crossover in the conduction at $T^*$ 
\cite{foot0}. 
The low-$T$ VRH conduction of the nonmetallic samples indicate that 
their DOS around $\mu$ is finite but the states therein are 
weakly localized [right column of Fig.\ \ref{fig3}(b) \cite{Molenda}], 
most likely due to the disorder potentials 
originating from the random distribution of the Li ions. 
We do not exclude the possibility of a first-order Mott transition 
occurring upon increasing $x >$ 0.87 \cite{Menetrier, Marianetti_NMat}, 
so that an energy gap opens around $\mu$ at higher Li concentration due to 
correlation. 

In contrast to Li$_x$CoO$_2$ at $x > x_c$ 
exhibiting a nonmetallic conduction, 
Na$_x$CoO$_2$ at high Na concentration 
exhibits a metallic conduction \cite{Lee_NMat}. 
Since the distance between the $A$-ion layer and the CoO$_2$ sheet 
is shorter in Li$_x$CoO$_2$ 
than in Na$_x$CoO$_2$ (Li/Na ions in Li$_x$CoO$_2$/Na$_x$CoO$_2$ 
are located in an octahedral/prismatic oxygen environment 
\cite{Mendiboure}), 
the low-energy excitations occurring in the CoO$_2$ sheets of Li$_x$CoO$_2$ 
may be more strongly affected by the Li-ion disorder 
to show weak localizations at $x > x_c$, 
whereas those in Na$_x$CoO$_2$ may be less affected by the disorder so that 
they can propagate through as coherent quasiparticles \cite{Valla, Hasan} 
to show a metallic conduction [lower left of Fig.\ \ref{fig3}(b)]. 
The effect of disorder of the $A$-ion layers on the transport properties 
was also demonstrated in 
Ca$_{0.33}$CoO$_2$ thin films which can adopt two types of 
Ca-ion arrangements: 
a defect-rich orthorhombic type and a defect-poor hexagonal 
type showing nonmetallic and metallic conductions, respectively 
\cite{Sugiura_APEX}.

Figure \ref{fig3}(a) shows $S$-$T$ curves of the Li$_x$CoO$_2$ 
epitaxial films. The positive value of $S$ in the whole 
temperature and composition range confirms that the 
carriers are holes in Li$_x$CoO$_2$. 
One can see that $S$ is increased with increasing $x$, 
and the $S$-$T$ curve of the $x =$ 0.87 sample exhibits a peak 
of magnitude $\sim$120 $\mu$V/K at $\sim$140 K. 
Note that this temperature coincides 
with $T^*$ determined from the $\rho$-$T$ analysis. 
In the polycrystalline bulk samples, the 140-K anomaly in the $S$-$T$ curves 
appears as a hump feature \cite{Molenda, Menetrier}. 
The increase in $S$ with $x$ and the peak in the $S$-$T$ curve 
at high Li concentration are similar to those 
observed in Na$_x$CoO$_2$ at high Na concentration 
\cite{Lee_NMat}, and therefore, these features 
are considered to be common for $A_x$CoO$_2$ at 
high $x$. 
Such a peak in the $S$-$T$ curves were also reported in the 
studies of carrier-doped transition metal compounds 
\cite{Zvyagin} such as Cu$_x$TiSe$_2$ \cite{CuTiSe2}, 
Fe$_3$O$_{4-x}$F$_x$ \cite{Fe3O4}, (CaOH)$_x$CoO$_2$ \cite{CaOHCoO2} 
and Fe$_x$ZrSe$_2$ \cite{FeZrSe}. 
On the other hand, the $S$-$T$ curves for the 
$x \le$ 0.75 samples are almost linear in $T$ up to 
$\sim$100 K, indicating the metallic conduction consistent with 
$\partial\rho$/$\partial T >$  0. 

The behavior of $S$ 
approaching to 0 with decreasing $T$ instead of 
diverging as $S \propto 1/T$ of an activated-type 
conduction \cite{Molenda}
further supports that the DOS of the nonmetallic 
samples is finite around $\mu$  
as shown in Fig.\ \ref{fig3}(b). 
When the conduction is a VRH and the DOS around 
$\mu$ is a slowly varying function of energy,
$S\propto T^{p}$, where $p = \frac{d-1}{d+1}$ ($\textless$ 1) \cite{Zvyagin}, 
so that the $S$-$T$ curves exhibit 
negative curvatures in contrast to 
$S\propto T$ (corresponding to $p = 1$) of a 
metallic conduction. 
We overlay on the 
$S$-$T$ curve of $x$ = 0.87 in Fig.\ \ref{fig3}(a) 
the theoretical relationship of the VRH 
for $d$ = 2 and 3 \cite{foot1}, and find that 
the negative curvature of the $S$-$T$ curve at $T \textless T^*$ is 
qualitatively 
reproduced by the theory. 
Thus, the analyses of $\rho$-$T$ and $S$-$T$ 
consistently indicate that, at least at $T<T^*$, the transport is 
carried by the low-energy excitations which are weakly 
localized due to disorder. 

Now we investigate the origin of the crossover in the transport at 
$T^* \sim$ 140 K. 
If the DOS is independent of $T$ (a rigid-band picture), 
$T^*$ can be understood as the onset temperature of 
the conduction carried by the 
delocalized states beyond the mobility edge 
[upper right of Fig.\ \ref{fig3}(b) \cite{foot2}]. 
However, a peak in the $S$-$T$ curve also occurs 
in a {\it metallic} Na$_x$CoO$_2$ at high $x$ \cite{Lee_NMat}, 
and if the origin is the same as that of Li$_x$CoO$_2$ at $x$ = 0.87, 
the crossover temperature $T^*$ should be explained other than the 
VRH scenario {\it within the rigid-band picture}, since this picture is only 
applicable to nonmetallic materials. In fact, 
there are signatures of non-rigid-band evolution of the 
electronic structures \cite{Zunger, Abbate, Kellerman, Ishida} and strong 
electron correlations \cite{vanElp, Menetrier, Marianetti_NMat} 
in Li$_x$CoO$_2$.

We recall evidence for a crossover 
in the low-energy electronic excitations with $T$ in the 
layered cobaltates: in the 
angle-resolved photoemission spectra of Na$_x$CoO$_2$ 
\cite{Valla, Hasan} and Bi-Pb-Co-O \cite{Valla}, 
the quasiparticle peak disappears above $T_{coh}\sim$ 200 K 
[left bottom of Fig.\ \ref{fig3}(b)], 
so that the Bloch-Boltzmann description of the transport becomes 
invalid at $T > T_{coh}$. 
This may result in the crossover seen in the 
transport properties, i.e., $T^* \sim T_{coh}$. 
In the case for Li$_x$CoO$_2$ at $x > x_c$ , 
the low-energy excitations in the coherent part are further affected by the 
disorder to show weak localizations 
[right bottom of Fig.\ \ref{fig3}(b)], and hence the 
VRH conduction at $T < T^*$. 
The smallness of the renormalized Fermi energy $\varepsilon_F^*\sim 100$ meV 
due to strong correlations \cite{SpinPolaron_PRL, Brouet, 
Marianetti_vacancy, DMFT_RMP} 
may be the origin of $T_{coh}$ as well as $T^*$ occurring 
at ambient temperatures \cite{Ishida_JPSJ}. 
The peak at $T^*$ in the $S$-$T$ curve can occur when crossovering 
into a high-$T$ incoherent conduction \cite{KoshibaePRL}, where the 
carriers are viewed as classical particles hopping among the 
Co sites. 
Alternatively, temperature dependence in the 
rate of quasiparticle scatterings 
with spin fluctuations that develops in a so-called pudding-mold band 
may explain the peak in the $S$-$T$ curves \cite{Kuroki_JPSJ}, 
and therefore, determination of the band structures of $A_x$CoO$_2$ 
at high $x$ is necessary. 

In summary, we investigated $S$ and 
$\rho$ of the Li$_x$CoO$_2$ epitaxial thin films 
which showed a two-phase coexistence 
at 0.75 $\le x \le$ 0.87 similar to that reported for bulk Li$_x$CoO$_2$. 
The resistivities were nearly 
an order of magnitude lower than those of the bulk 
polycrystals, presumably due to the reduction in the 
grain boundary effects in the epitaxial films. 
First, a metal-nonmetal transition occurred at $x \sim$ 0.8 in the 
biphasic domain, similar to those reported for bulk Li$_x$CoO$_2$. 
The nonmetallic conduction at high Li concentration implies that 
the Li-ion layers not only act 
as a charge reservoir for the CoO$_2$ sheets but also act as 
a source of disorder that strongly influence the 
conduction. Therefore, manipulating the 
strength of disorder through, e.g., changing the distance 
between the $A$-ion layer and the CoO$_2$ sheet 
would be a key to achieve high TE performance in $A_x$CoO$_2$. 
Second, we showed that the conduction of the 
nonmetallic samples exhibited an anomaly 
at $T^*\sim$ 140 K, where the $\rho$-$T$ curve deviated from 
the temperature dependence of the VRH and 
$S$ showed a drastic increase resulting into a peak 
in the $S$-$T$ curve at high Li concentration. 
We attributed $T^*$ to a crossover temperature 
from a low-$T$ VRH conduction to a high-$T$ incoherent conduction 
influenced by strong electron correlations. 

The authors acknowledge T.~Nonaka, Y.~Okamoto, 
and M.~Uchida for informative discussion.


\begin{thebibliography}{100}
\bibitem{foot}
Epitaxial strain is considered to be small since an 
amorphous Na-Al-O layer can be formed at the interface of the 
film and the substrate \cite{Mizutani}.
\bibitem{foot1}
If we use the relationship, $S = S_{corr} \propto T^p$, where $S_{corr}$ is a
  correlation contribution \cite{Limelette}, we get better a better fit to the
  data, although with an unphysically large negative value of $S_{corr}$.
  Therefore, we set $S_{corr}$ = 0.
\bibitem{foot2}
Within this picture, the $S$-$T$ curve of $x$ = 0.87 
at $T > T^*$ can be explained by an activated-type conduction 
such as a nearest neighbor hopping, $S-S_0\propto T^{-1}$, 
where $S_0$ is a constant; 
see Fig.\ \ref{fig3}(a) and Refs.\ \cite{CaOHCoO2, FeZrSe}.
\end{thebibliography}
\end{document}